\documentstyle[12pt,amssymb,epsf]{article}

\newtheorem{corollary}{Corollary}
\newtheorem{conjecture}{Conjecture}
\newtheorem{lemma}{Lemma}
\newtheorem{proposition}{Proposition}
\newtheorem{theorem}{Theorem}

\begin{document}

\begin{center}

{\Large \bf
Integrable Dynamics of Charges Related to Bilinear Hypergeometric
Equation
}\\*[3ex]

{\large \bf Igor Loutsenko}\\*[1ex]

SISSA, Via Beirut 2-4, 34014, Trieste, Italy \\
e-mail: loutseni@fm.sissa.it

\end{center}

\vspace{1.5cm}

\begin{center} \bf Abstract \end{center}

\begin{quote}

A family of systems related to a linear and bilinear evolution
of roots of polynomials in the complex plane is introduced.
Restricted to the line, the evolution induces dynamics of the
Coulomb charges (or point vortices) in
external potentials, while its fixed points correspond to
equilibriums of charges
in the plane. The construction reveals a direct connection with the
theories of
the Calogero-Moser systems and Lie-algebraic differential operators.
A study of the equilibrium configurations amounts in a construction
(bilinear hypergeometric equation) for
which the classical orthogonal and
the Adler-Moser polynomials represent some particular cases.

\end{quote}

\vspace{1cm}


\newpage

\section*{I. Introduction}

In the present paper we propose to discuss a {\it Bilinear 
Hypergeometric Operator}
\begin{equation}
\begin{array}{c}
H_\lambda^\Lambda[f,g]=
\left(f^{\prime\prime}g-2\Lambda f^\prime g^\prime+
\Lambda^2g^{\prime\prime}f\right)P+
\frac{1}{2}\left(f^\prime g+\Lambda^2g^\prime f\right)P^\prime
+\left(f^\prime g-\Lambda g^\prime f\right)U+\lambda fg\\
\\
P:=P(z)=A+B z+ C z^2,\quad U:=U(z)=a + b z,\quad 
\Lambda \in { \Bbb R } \\
\\
f:=f(z),\, g:=g(z),\quad f^\prime:=\frac{df(z)}{dz}
\end{array}
\label{bihop}
\end{equation}
and to study integrable dynamics:
\begin{equation}
\begin{array}{l}
\frac{d}{dt}x_i=2P(x_i)
\left(
-\sum_{j=1,j\not=i}^n\frac{1}{x_i-x_j}
+\sum_{j=1}^m\frac{\Lambda}{x_i-y_j}
\right)
-U(x_i)-\frac{1}{2}P^\prime(x_i)\\
\\
\frac{d}{dt}y_i=2P(y_i)
\left(
\sum_{j=1,i\not=i}^m\frac{\Lambda}{y_i-y_j}-
\sum_{j=1}^n\frac{1}{y_i-x_j}
\right)
-U(y_i)+\frac{\Lambda}{2}P^\prime(y_i)\\
\end{array}
\label{bidyn}
\end{equation}
of roots $x_i$, $y_i$ of polynomials in a complex variable $z$
$$
q_m(z,t)=\prod_{i=1}^m(z-x_i(t)), \quad
p_n(z,t)=\prod_{i=1}^n(z-y_i(t))
$$
induced by the action of (\ref{bihop}) 
\begin{equation}
p_n\frac{dq_m}{dt}-\Lambda q_m\frac{dp_n}{dt}=
H_{\lambda_{nm}}^\Lambda[p_n,q_m],\quad
\lambda_{nm}=(\Lambda m-n)
\left(U^\prime+(n-\Lambda m)P^{\prime\prime}/2\right)
\label{bieqs}
\end{equation}
The above construction has a nice physical interpretation:
The fixed points of (\ref{bidyn}) correspond to 
equilibrium distributions of $n$ and $m$ Coulomb
charges (or point vortices in hydrodynamics) of values 1 and 
$-\Lambda$ respectively in external potentials on
the plane or cylinder,
while the real solutions describe their motion on the
line or circle.

It should, however, be noted that the physical
analogies are not complete, because $dx_i^*/dt$ and $dy_i^*/dt$
should appear
in place of $dx_i/dt$ and $dy_i/dt$ in the lhs of (\ref{bidyn}). 
Nevertheless,
all equilibrium solutions in the plane
as well as time dependent solutions on the real line
are same for (\ref{bidyn}) and corresponding physical systems

The electrostatic interpretation for roots goes back to
works by Stieltjes on the classical orthogonal \cite{S},
and by Bartman on the Adler-Moser polynomials
\cite{Bartman}.
These results represent special cases in our construction.

The Bilinear Hypergeometric Equation
\begin{equation}
H_\lambda^\Lambda[f,g](z)=0
\label{bihyp}
\end{equation}
is a natural extension of
the Gauss hypergeometric equation
$$
H_\lambda^\Lambda[f,1](z)=
P(z)f^{\prime\prime}(z)
+\left(U(z)+\frac{1}{2}P^\prime(z)\right)f^\prime+\lambda f(z)=0
$$ 
and the recurrent relation
$$
H_0^1[\theta_i,\theta_{i+1}](z)
=\theta_{i+1}^{\prime\prime}(z)\theta_i(z)
-2\theta_{i}^\prime(z)\theta_{i+1}^\prime(z)
+\theta_{i+1}(z)\theta_i^{\prime\prime}(z)=0,
\quad
P=1,\, U=0
$$ 
for the Adler-Moser polynomials $\theta_i(z)$ \cite{AM}.
Another special case
$$
\Lambda=1,\, P(z)=-z^2,\, U=0 
$$
of (\ref{bihyp})
provides an interpretation for Huygensian
polynomials in two variables studied by Y.Berest and
the author in connection with the Hadamard
problem \cite{BeLou}.

The paper is organized as follows:

In the next two sections we study a linear evolution of polynomials,
which is a particular case of the bilinear dynamics.
We show that such an evolution
corresponds to dynamics of pairwise interacting charges iff it is
induced by second order
Lie-algebraic (hypergeometric or ``quasi exactly solvable'' )
differential operators.
The system of charges then can be embedded into an
integrable Hamiltonian
(in general, elliptic Calogero-Moser or Inozemtsev) model related 
to a Coxeter root system.
Skipping consideration of the ``quasi'' and elliptic cases
in the sequel , we classify remaining cases
by types of corresponding hypergeometric equations.

Sections IV-VI are devoted to the study of
the bilinear evolution and
its fixed points in the special case $\Lambda=1$:
Introducing the bilinear hypergeometric
operator, we show that its action induces dynamics, which,
in some settings, can be interpreted as a motion of unit
positive and negative Coulomb charges (or a system of vortices
in hydrodynamics) in external potentials.
We find that it can be embedded
in a flow generated by a sum of
two independent Calogero(Sutherland)-Moser Hamiltonians. In this way (\ref{bidyn}) can be integrated by the Lax method if 
$\Lambda=1$.

Considering polynomial solutions to $H_\lambda^1[f,g]=0$ in 
section V,
we analyze equilibrium configurations of charges.
It turns out that such solutions can be
obtained from associated linear problems by a finite
number of the Darboux transformations.

In section VI we discuss degenerate limits of the bilinear 
dynamics related
with the Kadomtsev-Petviashvilli equation and give an 
interpretation to a
set of algebraic solutions of a particular type of (\ref{bihyp}) 
obtained
earlier
in connection with the Hadamard problem in two dimensions.

In section VII we introduce polylinear ($l$-linear) hypergeometric 
operators and related dynamical systems, 
with (\ref{bihop})-(\ref{bieqs}) presenting a special case $l=2$. 
In this picture, $l$ distinct types of charges move in external 
potentials, interacting with each other. Such a dynamics can be 
embedded in a Hamiltonian system of $l$ species of particles of a 
Calogero-Moser type. In contrast with the $l=2,\Lambda=1$ case, 
there is no separation of the Hamiltonian flow in independent 
components and the Calogero(Sutherland)-Moser type potentials are
not related to the Coxeter root systems.

In section VIII we return to generic two component dynamics, presenting 
some arguments in favor of integrability of (\ref{bidyn}) for 
arbitrary real $\Lambda$.

Some open questions are discussed
in the concluding section of the paper.

\section*{II. Linear Evolution}

This section, which is a generalization of works by Choodnovsky
\& Choodnovsky \cite{CC} and Calogero \cite{Ca}, 
is devoted to the study of a particular case of the bilinear 
evolution.

Let
\begin{equation}
{\rm \bf V}={\rm Span}\{1,z,\dots,z^{n-1},z^n\}
\label{basis}
\end{equation}
be a linear space of polynomials over the complex
numbers ${\Bbb C}$, ${\rm \bf V} \cong {\Bbb C}^{n+1}$
of the degree less or equal to $n$ in $z$.

We consider the evolution of polynomials $p(z,t)$
\begin{equation}
\frac{dp(z,t)}{dt}=L[p(z,t)],
\quad p(z,t)=T(t)\prod_{i=1}^n (z-z_i(t)) \in {\rm \bf V}
\label{linear}
\end{equation}
under the action of a time independent linear operator
$L \in {\rm End({\bf V},{\bf V})}$.

Rewriting (\ref{linear}) in terms of roots
$z_i(t), i=1 \dots n$ and the common factor $T(t)$
we arrive to the following

\begin{lemma}
\label{rational}
The linear evolution equation (\ref{linear}) is equivalent to
the following dynamical system
\begin{equation}
\begin{array}{lr}
T^{-1}\frac{d}{dt}T=\tau(z_1 \dots z_n),&\qquad (a)\\
\\
\frac{d}{dt}z_i=v(z_i|z_1,z_2,\dots,{\hat z}_i,\dots,z_{n-1}, z_n),
\quad i=1\dots n, & \qquad (b)
\end{array}
\label{roots}
\end{equation}
$v$ is a rational
function symmetric in the last
$n-1$ (all but $z_i$, the hat in (\ref{roots}a) denotes omission of
$z_i$) variables.
\end{lemma}

Proof: Representing $L$ in the matrix form
\begin{equation}
L[z^i]=\sum_{j=0}^nL_{ij}z^j, \quad L_{ij}\in{\Bbb C}
\label{series}
\end{equation}
we equate the lhs and the rhs coefficients at different powers
of $z$ in (\ref{linear}). Equation (\ref{roots}a) is obtained by
picking out a coefficient at $z^0=1$.
Expressing $dT/dt$ from (\ref{roots}a) and substituting it to
the rest of equations we get:
\begin{equation}
\frac{d\sigma_i}{dt}=f_i(z_1...z_n), \quad i=1...n
\label{symmetric}
\end{equation}
where $f_i$ are polynomials symmetric in $z_i, i=1..n$ and $\sigma_i$
stand for elementary symmetric polynomials
$
\sigma_0(z_1 \dots z_n)=1,
\sigma_1(z_1 \dots z_n)=\sum_{i=1}^n z_i,
\sigma_2(z_1 \dots z_n)=\sum_{i<j} z_i z_j, \dots
$

Equation (\ref{symmetric}) is a linear system for $dz_i/dt$ with 
determinant $\prod_{i<j}(z_i-z_j)$ . It is not singular,
provided all $z_i$ are distinct, and (\ref{symmetric}) has a unique
solution. 

Since $f_i$ are symmetric in all arguments, this completes the proof.

\vspace{.1in}

We call (\ref{roots}) a system {\it with two body interactions} if
\begin{equation}
v(z_i|z_1,z_2,\dots,{\hat z}_i,\dots,z_{n-1}, z_n)
=\sum_{j\in N \not= i} w(z_i,z_j)+u(z_i)
\label{couples}
\end{equation}
In (\ref{couples}) and in the sequel the  following notations 
$N:=\{1,2,...n\}$, $M:=\{1,2,...m\}$ are used.

\begin{theorem}: 
\label{differen}
A system generated by (\ref{linear}) for $n>2$ is a system with 
the two-body
interaction if and only if $L$ is a
(modulo adding a constant) second order differential operator
\begin{equation}
\begin{array}{c}
L=P(z)\frac{d^2}{dz^2}
+U(z)\frac{d}{dz}
-\frac{n}{2}U^\prime(z)-\frac{n(n-1)}{6}P^{\prime\prime}(z)\\
\\
P(z)=A+Bz+Cz^2+Dz^3+Ez^4,\quad U(z)=a+bz+cz^2-2(n-1)Ez^3
\end{array}
\label{differential}
\end{equation}
with polynomial coefficients $P(z)$ and $U(z)$ at most degree four
and three respectively.

Under condition (\ref{differential}), (\ref{roots}) becomes
\begin{equation}
\frac{dz_i}{dt}=
-2P(z_i)\sum_{ j \in N \not=i }\frac{1}{z_i-z_j}-U(z_i),\\
\label{onecomponent}
\end{equation}
\end{theorem}

In other words, (\ref{onecomponent}) is the most general system
with two body interactions induced by a linear evolution
in the polynomial space.

{\bf Remark}: The cases $ n=1,2 $ are excluded from the theorem, 
since any
linear operator induces a two-body dynamics.

Proof: Introducing the following set of differential operators
$$
{\cal L}_{ij}=\frac{(-)^iz^j}{(n-i)!}
\prod_{k=1}^{n-i}\left(z\frac{d}{dz}-k\right)\frac{d^i}{dz^i},\quad
{\cal L}_{ij}[z^k]=\delta_{ik}z^j,\quad 0\leq i,j,k\leq n
$$
we see from (\ref{series}), that any $L\in {\rm End}({\rm\bf V},
{\rm\bf V})$
can be represented by a differential operator with
polynomial coefficients $Q_i(z)$
\begin{equation}
L=\sum_{i=0}^n\sum_{j=0}^nL_{ij}{\cal L}_{ij}
=\sum_{i=0}^nQ_i(z)\frac{d^i}{dz^i}
\label{temp}
\end{equation}

Substituting (\ref{temp}) to (\ref{linear}), using (\ref{roots}a) and
imposing condition (\ref{couples}) we have:
$$
\begin{array}{l}
Q:=\tau(z_1,z_2,\dots)-Q_0(z)-\sum_i\left(Q_1(z)-u(z_i)\right)t_i\\
\\
\qquad\qquad
-2\sum_{i<j}\left(Q_2(z)-w(z_i,z_j)\right)t_it_j
-6\sum_{i<j<k}Q_3(z)t_it_jt_k-.....=0
\end{array}
$$
where $t_i=1/(z-z_i)$.

Taking the $n+1$ th partial derivative
$
\frac{\partial^{n+1}Q}{\partial z\partial z_1\partial z_2
\dots \partial z_{n-1}\partial z_n}=0
$
we get the following equation
$$
(t_1...t_n)^2\left(2Q_n(z)\sum_{ i \in N }t_i-Q_n^\prime(z)\right)=0
$$
Picking out the coefficients at different powers of $t_i, i\in N$,
we find that $Q_n(z)=0$,
since $z,t_1....t_n$ is a set of independent variables.

Proceeding by induction and taking
the $n, n-1, ...4$th derivative of $Q$
$$
\frac{\partial^nQ}{\partial z\partial z_{i_1}\partial z_{i_2}
\dots\partial z_{i_j}}=0, \quad i_1<i_2<...<i_j,\quad 2<j<n
$$
we eliminate all $Q_i$ with $i>2$.

Hence , the operator $L$ is at most of the second order.

Decomposing $L$ into the sum of homogeneous components
$$
L=\sum_i L_i,\quad L_i: {\Bbb C}z^j \to {\Bbb C}z^{i+j},
\quad L_i=a_iz^{i+2}\frac{d^2}{dz^2}+b_iz^{i+1}\frac{d}{dz}+c_iz^i
$$
and remembering that ${\rm\bf V}$ is spanned
by $z^i, i=0..n$ only, we have
$$
L_i[z^{n-2}]=L_i[z^{n-1}]=L_i[z^n]=0, \quad {\rm for} \quad i>2
$$
It is immediate that
$$
a_i=b_i=c_i=0, \quad {\rm for} \quad i>2
$$
Proceeding in this way, we impose similar conditions for $i=1$ and 
$i=2$
$$
L_2[z^{n-1}]=L_2[z^n]=L_1[z^n]=0
$$
getting the most general expression (\ref{differential}) for a
second order differential operator
$L\in {\rm End}({\rm \bf V},{\rm \bf V})$.

The sufficient condition of the theorem is proved by
the direct calculation: Substituting (\ref{onecomponent}) and
$$
T^{-1}\frac{d}{dt}T=
L_{nn}+L_{n-1,n}\sum_{i=1}^nz_i+L_{n-2,n}\sum_{i<j}z_iz_j
$$
into (\ref{differential}), expressing the matrix elements $L_{ij}$
in terms of $ a,b,c,A,B,C,D,E,n$ and using the following identity
\begin{equation}
\frac{1}{(z-z_i)(z-z_j)}=\frac{1}{z_i-z_j}
\left(\frac{1}{z-z_i}-\frac{1}{z-z_j}\right)
\label{simple}
\end{equation}
we show that (\ref{linear}) holds identically. This completes the 
proof.

\vspace{5mm}

Although (\ref{onecomponent}) is not a Hamiltonian system,
the following proposition holds

\begin{proposition}
\label{hamilton}
(\ref{onecomponent}) is a trajectory of a system with the Hamiltonian
\begin{equation}
\begin{array}{c}
H=\sum_{ i \in N } \frac{1}{2P(z_i)}\left(\frac{dz_i}{dt}\right)^2-V, \\
\\
V=2\sum_{i<j}\frac{P(z_i)+P(z_j)}{(z_i-z_j)^2}
+(n-2)\left(E(z_i+z_j)^2+D(z_i+z_j)\right)
+\frac{U(z_i)-U(z_j)}{z_i-z_j}
+\frac{1}{2}\sum_i\frac{U^2(z_i)}{P(z_i)}
\end{array}
\label{hamiltoni}
\end{equation}

In other words, the Hamiltonian equations of motion
\begin{equation}
\frac{1}{P(z_i)^2}\left(\frac{d^2z_i}{dt^2}\right)
-\frac{P^\prime(z_i)}{2P(z_i)^2}\left(\frac{dz_i}{dt}\right)^2
=\frac{\partial V}{\partial z_i},\quad i \in N
\label{acceleration}
\end{equation}
are corollaries of (\ref{onecomponent}).
\end{proposition}
{\bf Remark}: Equations (\ref{acceleration}) are reduced to
the Newtonian form
\begin{equation}
\frac{d^2\phi_i}{dt^2}=\frac{\partial V}{\partial\phi_i}, 
\quad i \in N
\label{newvar}
\end{equation}
by the change of variables
\begin{equation}
\phi_i=\zeta(z_i),\quad \frac{d\zeta(z)}{dz}=\frac{1}{\sqrt{P(z)}} 
\label{new}
\end{equation}

\vspace{7.5mm}

Proof of proposition \ref{hamilton}: Expressing second derivatives 
through (\ref{onecomponent})
$\frac{d^2z_i}{dt^2}=
\sum_{j\in N}\frac{dz_j}{dt}\frac{\partial}{\partial z_j}
\left(\frac{dz_i}{dt}\right)$
by direct calculations, we evaluate the lhs of (\ref{acceleration}):
$$
{\rm lhs(\ref{acceleration})}=-\frac{\partial}{\partial z_i}\left(
2\sum_{i,n\not=i,j\not=i}\frac{P(z_i)}{(z_j-z_i)(z_n-z_i)}
+\sum_{i\not=j}\frac{U(z_i)-U(z_j)}{z_i-z_j}
+\frac{1}{2}\sum_i\frac{U^2(z_i)}{P(z_i)}\right)
$$
Using the identity
$$
\sum_{i,n\not=i,j\not=i}\frac{P(z_i)}{(z_j-z_i)(z_n-z_i)}=
\sum_{i<j}\frac{P(z_i)+P(z_j)}{(z_i-z_j)^2}
+(n-2)\left(E(z_i+z_j)^2+D(z_i+z_j)+\frac{2}{3}C\right)
$$
we get (\ref{hamiltoni}), which competes the proof.

\section*{III. System Classification, Lie-Algebraic Operators and\\
Calogero$-$Moser Models}

Consider the $n+1$ dimensional representation of the Lie algebra 
$sl(2,{\Bbb C})$
$$
\left[J^0,J^\pm\right]=\pm J^\pm, \quad\left[J^-,J^+\right]=2J^0
$$
by differential operators
\begin{equation}
J^+=z^2\frac{d}{dz}-nz,\quad J^0=z\frac{d}{dz}-\frac{n}{2},
\quad J^-=\frac{d}{dz}
\label{sl}
\end{equation}
acting on (\ref{basis}).

Operator (\ref{differential}) is an element of the universal 
enveloping
algebra of $sl(2,{\Bbb C})$, $L=\sum\alpha_{\upsilon\varsigma}
J^\upsilon J^\varsigma+\beta_\varsigma J^\varsigma, 
\upsilon,\varsigma=\pm,0$
(\ref{sl}). Such Lie-algebraic operators are called 
``quasi-exactly solvable''
\cite{Turbin}.
They can be separated in nine nontrivial equivalence classes
under the linear-fractional transformations of the independent 
variable $z$.
Based on the invariant-theoretic
classification of canonical forms for quartic polynomials 
\cite{Gurevich}, operators
(\ref{differential})
can be placed in the nine canonical forms with
\begin{equation}
\begin{array}{llll}
(1) & P(z)=1 &\quad (4,5) & P(z)=1\mp z^2      \\
(2) & P(z)=z &\quad (6,7) & P(z)=(1\mp z^2)^2  \\
(3) & P(z)=-z^2 &\quad (8,9) & P(z)=(z^4+\tau z^2\mp 1)
\end{array}
\label{list}
\end{equation}
Most general Hamiltonian systems (\ref{hamiltoni}) in this 
classification
are elliptic Inozemtsev models (for trigonometric
and rational Inozemtsev Modes see e.g. \cite{I}, \cite{P} ) 
related to ${\rm A_n}$ and
${\rm BC/D_n}$ Coxeter root systems.

Skipping analysis of ``quasi'' and elliptic cases in the sequel, 
we content ourselves with ``exactly solvable'' 
hypergeometric operators
dealing with the first five classes only (and linear $U(z)=a+bz$) in
(\ref{list}).

Figure \ref{table} gathers needed information on the first four 
(the fifth one is a hyperbolic
version of the fourth) classes, which are related to the 
rational/trigonometric Calogero(Sutherland)-Moser systems.

\begin{figure}
\begin{tabular}{||c|c|c|c|c||} \hline
$ P(z)        $ & $ z=z(\phi)   $ &   Root   & Calogero-Moser                      &      Polynomial Eigen-      \\
$             $ & $ (\ref{new}) $ &  System  & Potential $V(z(\phi))$, (\ref{hamiltoni}), (\ref{newvar}) & functions of $L$ (\ref{differential})\\ \hline

$  1   $ & $  z=\phi   $ & $ {\rm A}_n  $ &
$ \sum_{j<k}\frac{4}{(\phi_k-\phi_j)^2}+\frac{1}{2}\sum_{j}\left(a+b\phi_j\right)^2 $ & $ \begin{array}{c} {\rm if}\, b\not = 0 \, {\rm Hermite} \\ H_n\left(\sqrt{\frac{-b}{2}}(z+\frac{a}{b})\right)\\ {\rm if}\, b=0,\\ 1,z \end{array} $ \\ \hline

$  z  $ & $z=\frac{\phi^2}{4} $ & ${\rm BC/D}_n$ & $ \begin{array}{r} \sum_{j<k}\frac{4}{(\phi_k-\phi_j)^2}+\frac{4}{(\phi_k+\phi_j)^2}\\+\sum_{j}\frac{2a^2}{\phi_j^2}+\frac{b^2\phi^2_j}{8} \end{array} $  & $ \begin{array}{c} {\rm if}\, b\not=0\, {\rm Laguerre} \\ L_n^{a-1}(-bz)\\ {\rm if}\,b=0,a=1-j \\ 1,z^j \end{array} $ \\ \hline

$ -z^2 $ & $ z=\exp(i\phi) $ &$ {\rm A}_n$& $ \sum_{j<k}\frac{1}{\sin^2(\phi_k-\phi_j)/2}+\frac{1}{2}\sum_{j}ae^{-i\phi_j}+be^{i\phi_j} $ & $ \begin{array}{c} \\ z^nL^{b-2n+1}_n\left(\frac{-a}{z}\right) \\ \mu(z)=z^{b-2n+1}e^\frac{a}{z} \end{array} $ \\ \hline

$ 1-z^2  $ & $ z=\cos(\phi) $ &$ {\rm BC/D}_n$& $\begin{array}{r} \sum_{j<k}\frac{1}{\sin^2(\phi_k-\phi_j)/2}+\frac{1}{\sin^2(\phi_k+\phi_j)/2}\\+\sum_{j}\frac{(b+a)^2}{2\sin^2\phi_j}-\frac{ab}{2\cos^2\phi_j/2} \end{array} $ & $\begin{array}{c} \\ {\rm Jacobi}\\ P_n^{(-\frac{a+b+2}{2},\frac{a-2-b}{2})}(z) \\ \, \end{array} $ \\ \hline

\end{tabular}
\caption{Four generic classes of Hypergeometric systems}
\label{table}
\end{figure}

\section*{IV. Bilinear Evolution, $\Lambda=1$}

In this section we introduce a special case 
$H_\lambda[\cdot,\cdot]:=H^1_\lambda[\cdot, \cdot]$ 
of the  Bilinear Hypergeometric Operator (\ref{bihop})
and study related integrable dynamics of roots.

Let
$$
{\rm\bf V}_1 \cong {\Bbb C}^{n+1}, {\rm\bf V}_2 \cong {\Bbb C}^{m+1},
{\rm\bf V}_3 \cong {\Bbb C}^{m+n}
$$
be linear spaces of polynomials of degree less or equal to $n$, $m$ and
$n+m-1$ respectively.

Consider the evolution
\begin{equation}
q\frac{dp}{dt}-p\frac{dq}{dt}=H_\lambda[p,q],\quad p\in {\rm\bf V}_1, 
q \in {\rm\bf V}_2 
\label{bilinear}
\end{equation}
under the action of a bilinear operator $ H_\lambda: V_1\times V_2 \to V_3 $
on the monic polynomials
\begin{equation}
p=\prod_{ i \in N }(z-x_i(t)),\quad q=\prod_{ i \in M }(z-y_i(t))
\label{polyn}
\end{equation}
of the $n$th and $m$th degrees.

\begin{lemma}
\label{rationals}
The bilinear evolution equation (\ref{bilinear}) is equivalent to
the following dynamical system
$$
\begin{array}{lr}
\frac{d}{dt}x_i=v_1(x_i|x_1..{\hat x}_i..x_n|y_1..y_m),\quad i=1\dots n\\
\\
\frac{d}{dt}y_i=v_2(y_i|y_1..{\hat y}_i..y_m|x_1..x_n),\quad i=1\dots m
\label{twocomponn}
\end{array}
$$
where $v_1$ and $v_2$ are rational functions ($v_1$ is symmetric in 
$x_1..{\hat x}_i..x_n$ and $y$
. $v_2$ is symmetric in $y_1..{\hat y}_i..y_m$ and $x$).
\end{lemma}

Proof: is similar to the linear case (lemma \ref{rational}), 
except that the polynomials are monic now. $dx/dt$ and $dy/dt$ 
are uniquely expressed from a
linear system of equations with the determinant 
$\prod_{j<i\in N}(x_j-x_i)\prod_{j\in N, i\in M}(x_j-y_i)
\prod_{i<j\in M}(y_i-y_j)$. It is non singular
provided all roots are distinct.

Again, we study systems with two body interactions which
means that 
$$
\begin{array}{l}
v_1(x_i|\hat x|y)=\sum_{j\not=i} w_{11}(x_i,x_j)+\sum_{i,j} w_{12}(x_i,y_j) \\
\\
v_2(y_i|\hat y|x)=\sum_{j\not=i} w_{22}(y_i,y_j)+\sum_{i,j} w_{21}(y_i,x_j)
\end{array}
$$

As in the linear case, it is natural to look for differential operators
inducing integrable dynamics in a system with two-body interactions. It turns
out that such operators exist and are extensions of the linear
case.

\begin{proposition}
\label{bilinears}
The bilinear operator  $H_{\lambda_{nm}}: V_1\times V_2 \to V_3 $
\begin{equation}
H_{\lambda_{nm}}\left[p,q\right]=(p^{\prime\prime}q-2p^\prime q^\prime+pq^{\prime\prime})P
+\frac{1}{2}(p^\prime q+pq^\prime)P^\prime+(p^\prime q-q^\prime p)U+
\lambda_{nm}pq
\label{smile}
\end{equation}
$$
P(z)=A+Bz+Cz^2,\quad U(z)=a+bz,
$$
$$
\lambda_{nm}=(m-n)\left(U^\prime+(n-m)P^{\prime\prime}/2\right)
$$
induces dynamics
\begin{equation}
\begin{array}{l}
\frac{d}{dt}x_i=2P(x_i)
\left(
-\sum_{j=1,j\not=i}^n\frac{1}{x_i-x_j}+\sum_{j=1}^m\frac{1}{x_i-y_j}
\right)
-U(x_i)-\frac{1}{2}P^\prime(x_i)\\
\\
\frac{d}{dt}y_i=2P(y_i)
\left(
\sum_{j=1,i\not=i}^m\frac{1}{y_i-y_j}-\sum_{j=1}^n\frac{1}{y_i-x_j}
\right)
-U(y_i)+\frac{1}{2}P^\prime(y_i)\\
\end{array}
\label{biun}
\end{equation}
by action (\ref{bilinear}) on (\ref{polyn}).
\end{proposition}

{\bf Remark}: Equation (\ref{bilinear}), (\ref{smile}) may be 
written in the form of the Schr\"odinger evolution equation
with a time-dependent potential 
\begin{equation}
\frac{d\psi}{dt}=P(z)\psi^{\prime\prime}+(U+1/2P^\prime)\psi^\prime+(P^\prime(\ln q)^\prime-2P(\ln q)^{\prime\prime}+\lambda)\psi
\label{lianerize}
\end{equation}
where $\psi=p/q$. In this setting, we study the time evolution of
polynomial $q$ 
and a rational function $\psi$ with denominator $q$, which is
a rather inconvenient formulation for our purposes.

{\bf Remark}: Substituting $P(z)=1$, $U(z)=-k, (dk/dz=0)$, and 
$\lambda=0$  in (\ref{lianerize})  and reexpressing it in the 
formally self-adjoint form
we obtain the nonstationary Scr\"odinger equation
\begin{equation}
\frac{d\Phi}{dt}=\Phi^{\prime\prime}+{\cal U}\Phi,\, 
{\cal U}=-2(\log \tau)^{\prime\prime},  
\label{solitons}
\end{equation}
$$
\tau=q
$$
which is a second equation of an auxiliary linear problem for 
Kadomtsev-Petviasvilly hierarchy \cite{KP} (with $q$ as a 
$\tau$-function).
The solution to (\ref{solitons}) is now a quasirational function
$$
\Phi=\frac{p}{q}\exp(kz+k^2t)
$$
We observe similarities with Krichever construction \cite{Surfs} , 
\cite{Poles}  for the rational Baker-Achieser function.
In more details, the Backer-Akhieser function $\Psi(z,t,k)$ is a 
special $n=\deg(p(z))=m=\deg(q(z))$
case of the above quasirational function
$$
\Phi=\Psi=\left(1+\sum_{i=1}^n\frac{\eta_i(t,k)}{z-x_i(t)}\right)
\exp\left(kz+k^2t\right)
$$
with divisor of simple poles defined at points $x_i, i\in N, N=M$.

\vspace{7.8mm}

Proof of the proposition \ref{bilinears}: Is essentially similar 
to the proof of sufficient condition of
theorem \ref{differen}: one substitutes (\ref{biun}) in (\ref{smile})
and uses identity (\ref{simple}).

\vspace{7.8mm}

Similarly to the linear case, equations of motion (\ref{biun}) can be 
expressed
in the Newtonian coordinates (\ref{new})
$$
\frac{d\phi_i}{dt}=-\frac{\partial{\cal H}}{\partial\phi_i},
\quad\frac{d\theta_j}{dt}=\frac{\partial{\cal H}}{\partial\theta_j},\quad x_i=\zeta(\phi_i), y_j=\zeta(\theta_j)\, i \in N, \, j \in M
$$
\begin{equation}
{\cal H}=\sum_{ i<j \in N }w(\phi_i,\phi_j)+\sum_{ i \in N }u(\phi_i)-\sum_{ i \in N , j \in M }w(\phi_i,\theta_j)-\sum_{ i \in M } u(\theta_i )+\sum_{ i<j \in M }w(\theta_i,\theta_j)
\label{fields}
\end{equation}
\begin{equation}
w(\phi,\theta)=\ln\frac{\left(\zeta(\phi)-\zeta(\theta)\right)^2}{P(\zeta(\phi))P(\zeta(\theta))}, \quad P(z)\frac{du(\phi(z))}{dz}=U(z)-\frac{m-n}{2}P^\prime(z),
\label{interaction}
\end{equation}

Should we have $d\phi^*_i/dt$, $d\theta^*_i/dt$ instead of 
$d\phi_i/dt$ and $d\theta_i/dt$ in the
lhs of the equations of motion, the system (\ref{fields}) 
would be a Hamiltonian system of $n$ positive and $m$ negative
vortices or Coulomb charges on the plane or cylinder \cite{A}, \cite{AN}. 
It is not Hamiltonian in our case, but has the
same fixed points in the plane or cylinder and dynamics on the real 
line or circle.

Let us discuss the question of integrability of (\ref{biun}).

In the linear case, which is a particular case $m=0$ of
(\ref{bilinear}), there were two lines of approach to the 
integration of system
(\ref{onecomponent}):

For the first possibility, the linear dynamics (\ref{linear})
in the finite basis
(\ref{basis}) allowed us to find $p(z,t)$ (and $z_i(t)$) at any $t$.

The second way to find $z_i$ was to solve the Hamiltonian system
(\ref{newvar})
with initial conditions given by (\ref{onecomponent}) itself.

Obviously, the first of above approaches does not apply in the bilinear
case, since the evolution is not linear any more.
Therefore, we use the second method, trying to embed (\ref{biun})
into a Hamiltonian system.

\begin{lemma}
\label{identity}
If the odd function $\Phi(x)$ satisfies the functional equation
\begin{equation}
\Phi(x)\Phi(y)+\Phi(z)\Phi(x)+\Phi(y)\Phi(z)=0
\label{condition}
\end{equation}
whenever $x+y+z=0$. Then the following identities hold
\begin{equation}
I_1(x)=\sum_{i\in N, i\not=n}\sum_{j\in N, j\not=n}\sum_{n\in N}\Phi(x_n-x_i)\Phi(x_n-x_j)=2\sum_{i<j\in N}\Phi(x_i-x_j)^2
\label{identity1}
\end{equation}
$$
I_2(x,y)=2\sum_{m\in M}\sum_{i\in N}\sum_{j\in N,j\not=i}\Phi(x_j-y_m)\Phi(x_i-x_j)-
2\sum_{m\in N}\sum_{i\in M}\sum_{j\in M,j\not=i}\Phi(y_j-x_m)\Phi(y_i-y_j)
$$
\begin{equation}
+\sum_{m\in N}\sum_{i\in N}\sum_{j\in M}\Phi(y_j-x_m)\Phi(x_i-y_j)-
\sum_{m\in M}\sum_{i\in M}\sum_{j\in N}\Phi(x_j-y_m)\Phi(y_i-x_j)=0
\label{identity2}
\end{equation}
\end{lemma}

Proof: is a calculation.

\begin{theorem}
\label{hamiltonii}
(\ref{biun}) can be embedded into the flow generated by the sum of two
independent Hamiltonians
$$
H=H_++H_-,
$$
\begin{equation}
H_+=\sum_{ i \in N }\frac{1}{2P(x_i)}\left(\frac{dx_i}{dt}\right)^2-V_+(x), \\
\quad
H_-=-\sum_{ i \in M }\frac{1}{2P(y_i)}\left(\frac{dy_i}{dt}\right)^2+V_-(y) \\
\label{direct}
\end{equation}
$$
V_\pm(z)=2\sum_{i<j}\frac{P(z_i)+P(z_j)}{(z_i-z_j)^2}
+\frac{1}{2}\sum_i\frac{U_{\pm}^2(z_i)}{P(z_i)},
\quad
U_\pm(z)=U(z)\pm\frac{1}{2}P^\prime(z)
$$

In other words , the Hamiltonian equations of motion
\begin{equation}
\frac{d^2}{dt^2}\phi_i=\frac{\partial V_+(\zeta(\phi))}{\partial\phi_i},\quad
\frac{d^2}{dt^2}\theta_i=\frac{\partial V_-(\zeta(\theta))}{\partial\theta_i}
\label{hami}
\end{equation}
where $\zeta$ is given by (\ref{new}), are corollaries of (\ref{biun}).
\end{theorem}

{\bf Remark}: Some results related to the special case $P(z)=1$ 
of theorem \ref{hamiltonii} (rational $A_n$ root system) were 
obtained by Veselov \cite{EN}, who studied rational solutions of 
the Kadomtsev-Petviashvili equation. In particular, 
it was found that poles of (unbounded at infinity) rational 
solutions of the KP equation 
(which are coordinates $x_i, i \in N $ in our case) move under 
the Calogero-Moser flow with nonzero external potential. 
It is interesting to note that the nondegenerate external potentials 
$U_+$ and $U_-$ coincide only in the above mentioned special case.

\vspace{7.8mm}

Proof of theorem \ref{hamiltonii}: Let us check that the 
Hamiltonian equation of motion for $x_i$
\begin{equation}
\frac{1}{P(x_i)^2}\left(\frac{d^2x_i}{dt^2}\right)
-\frac{P^\prime(x_i)}{2P(x_i)^2}\left(\frac{dx_i}{dt}\right)^2
=\frac{\partial}{\partial x_i}V_+
\label{acceleration1}
\end{equation} 
holds, expressing the second derivatives through (\ref{bidyn})
$$
\frac{d^2x_i}{dt^2}=\sum_{j=1}^n\frac{dx_j}{dt}\frac{\partial}{\partial x_j}\left(\frac{dx_i}{dt}\right)+\sum_{j=1}^m\frac{dy_j}{dt}\frac{\partial}{\partial y_j}\left(\frac{dx_i}{dt}\right)
$$
By direct calculations we get
$$
{\rm lhs(\ref{acceleration1})}=-\frac{\partial}{\partial x_i}\left(2W_1(A,B,C|x,y)+4W_2(A,B,C|x)+\frac{1}{2}\frac{U_+(x_i)^2}{P(x_i)}\right)
$$
where
$$
W_1(A,B,C|x,y)=2\sum_{m\in M}\sum_{i\in N}\sum_{j\in N,j\not=i}\frac{P(x_j)}{(x_j-y_m)(x_i-x_j)}-
2\sum_{m\in N}\sum_{i\in M}\sum_{j\in M,j\not=i}\frac{P(y_j)}{(y_j-x_m)(y_i-y_j)}
$$
$$
+\sum_{m\in N}\sum_{i\in N}\sum_{j\in M}\frac{P(y_j)}{(y_j-x_m)(x_i-y_j)}-
\sum_{m\in M}\sum_{i\in M}\sum_{j\in N}\frac{P(x_j)}{(x_j-y_m)(y_i-x_j)}+
\sum_{i\in M}\sum_{j\in N}\frac{P^\prime(y_i)}{y_i-x_j}
$$
$$
W_2(A,B,C|x)=\sum_{i\in N, i\not=n}\sum_{j\in N, j\not=n}\sum_{n\in N}\frac{P(x_n)}{(x_n-x_i)(x_n-x_j)}
$$
and $ P(z)=A+Bz+Cz^2 $.

Let us evaluate $W_1(A,B,C|x,y)=AW_1(1,0,0|x,y)+BW_1(0,1,0|x,y)+CW_1(0,0,1|x,y)$.

In $W_1(1,0,0|x,y)$ we immediately recognize 
identity (\ref{identity2})
with $\Phi(x)=1/x$. Consequently
$$
W_1(1,0,0|x,y)=I_2(x,y)=0
$$
Changing variables $x_i=\exp(\phi_i)$, $y_i=\exp(\theta_i)$
we find that
$$
W_1(0,0,1|x,y)=I_2(\phi,\theta)+Anm(1-n+m)=Anm(1-n+m),\quad\Phi(x)=\coth(x).
$$
Finally, using linearity of $W_1$ with respect to parameters 
$A,B,C$ we write
$ W_1(0,1,0|x,y)=W_1(0,0,1|x+\frac{1}{2},y+\frac{1}{2})-W_1(0,0,1|x,y)-\frac{1}{4}W_1(0,0,1|x,y)=0 $

Therefore
$$
W_1(A,B,C|x,y)=Anm(1-n+m)
$$
Applying (\ref{identity1}), we evaluate $W_2$ in a similar 
way, getting (\ref{acceleration1}) with $V_+$ given in 
(\ref{direct}).

The proof is completed by applying a similar procedure to $y_i$.

\begin{corollary}
\label{only}
(\ref{only}) is integrated by the Lax method
\end{corollary}

Proof: Since (\ref{direct}) are Calogero(Sutherland)-Moser Hamiltonians,
equations of motions (\ref{hami}) can be represented in the Lax form \cite{P}.
Thus solutions
to (\ref{biun}) may be found from (\ref{direct}) subject to initial conditions
given by (\ref{biun}) itself.

\section*{V. Equilibrium Configurations, Bilinear Hypergeometric Equation}

The fixed points $dx_i/dt=0$ and $dy_i/dt=0$ of (\ref{biun}) 
describe equilibrium
of the unit positive and negative Coulomb charges in two dimensional
electrostatic or point vortices in
hydrodinamics respectively \cite{S}, \cite {AN}. 
The polynomials $p$ and $q$ (\ref{polyn})
must then satisfy an ordinary bilinear differential equation
\begin{equation}
H_{\lambda_{nm}}[p,q]=0
\label{hypergeometric}
\end{equation}
which is a special case 
$\Lambda=1$ of the Bilinear Hypergeometric Equation (\ref{bihyp}).

Studying the dynamics of roots we supposed that they are distinct
and polynomials $p$ and $q$ do not have common factors 
(lemmas \ref{rational}, \ref{rationals}). 
However, solutions of (\ref{hypergeometric}) may have multiple
roots or/and common factors. In this circumstances we need to modify 
(\ref{biun}).

{\bf Remark}: one does not encounter such a problem 
in the linear case since polynomial
solutions of ordinary Hypergeometric equation 
(classical orthogonal polynomials) do
not have multiple roots.

\begin{proposition}\label{multi}
Let $p$ and $q$ be polynomials of orders
$n$ and $m$ satisfying (\ref{hypergeometric}) and
$$
p/q={\bar p}/{\bar q}, \quad {\bar p}=\prod_{i=1}^{\bar n}(z-x_i)^{\nu_i}, \quad {\bar q}=\prod_{i=1}^{\bar m}(z-y_i)^{\sigma_i}
$$
where ${\bar p}$ and ${\bar q}$ do not have common roots. 
Then $x$ and $y$ are critical points
of the Energy function
\begin{equation}
{\cal H}(x,y)=\sum_{i<j=1}^{i={\bar n}, j={\bar n}}\nu_j\nu_iw(x_i,x_j)+\sum_{i=1}^{\bar n}\nu_iu(x_i)-\sum_{i,j=1}^{i={\bar n}, j={\bar m}}\nu_i\sigma_jw(y_i,x_j)+\sum_{i<j=1}^{i={\bar m}, j={\bar m}}\sigma_i\sigma_jw(y_i,y_j)-\sum_{j=1}^{\bar m}\sigma_j u(y_j)
\label{extremum}
\end{equation}
$$
w(x_1,x_2)=\ln\frac{(x_1-x_2)^2}{P(x_1)P(x_2)}
$$
\end{proposition}

In other words, the total charge at point $x_i$ equals
to the difference of multiplicities of the corresponding root in $p$ and $q$. 

Proof: It is straightforward to check that
$$
{\rm res}_{z=x_i}\frac{H_\lambda[p,q](z)}{p(z)q(z)}={\rm res}_{z=x_i}\frac{H_\lambda[{\bar p},{\bar q}](z)}{{\bar q}(z){\bar p}(z)}
$$
where $ {\rm res}_{z=x_i} $ stands for the residue of a 
simple pole in the point $x_i$.
The residue is zero since $p$ and $q$ satisfy (\ref{hypergeometric}).

By direct calculation we get
$$0={\rm res}_{z=x_i}\frac{H_\lambda[{\bar p},{\bar q}](z)}{{\bar p}(z){\bar q}(z)}=
\sum_{j=1,j\not=i}^{\bar n}\frac{2\nu_jP(x_i)}{x_i-x_j}+\sum_{j=1}^{\bar m}\frac{2\sigma_jP(x_i)}{x_i-y_j}
-\nu_iU(x_i)+\frac{1-2\nu_i}{2}P^\prime(x_i)
$$
which is a derivative $\partial{\cal H}/\partial x_i$ of the 
energy (\ref{extremum}). Repeating similar calculation for
$y_i$ we complete the proof.

The following proposition gives examples of 
equilibrium configurations
corresponding to several generic cases of the figure \ref{table}.

\begin{proposition}
\label{equilibrium}
Let $I=i_1<i_2 ... i_k<i_{k+1}$ be a strictly increasing sequence of
nonegative integers and let $Q_i(z)$ be classical orthogonal 
polynomials satisfying
the hypergeometric equation
\begin{equation}
(L+\lambda_i)Q_i(z)=0,\quad L=P(z)\frac{d^2}{dz^2}+U(z)\frac{d}{dz}
\label{stuli}
\end{equation}
where (up to a linear transformation of $z$)
$$
\begin{array}{lll}
{\rm (i)} & \quad P(z)=1 &\quad U(z)=bz \\
{\rm (ii)}& \quad P(z)=-z^2 &\quad U(z)=bz
\end{array}\qquad b \not = 0
$$
Then polynomials $p$ and $q$
\begin{equation}
\begin{array}{ll}
p(z)=P(z)^{\frac{1}{4}k(k+1)}{\cal W}[Q_{i_1}(z),Q_{i_2}(z), ...Q_{i_k}(z),Q_{i_{k+1}}(z)],\\
\\
q(z)=P(z)^{\frac{1}{4}(k-1)k}{\cal W}[Q_{i_1}(z),Q_{i_2}(z), ...Q_{i_k}(z)] 
\end{array}
\label{darboux}
\end{equation}
$$
\begin{array}{ll}
{\rm (i)} &\quad\deg(p)=n=\sum_{j=1}^{k+1} i_j-\frac{1}{2}k(k+1), \, \deg(q)=m=\sum_{j=1}^{k} i_j-\frac{1}{2}k(k-1)\\ \\
{\rm (ii)}&\quad\deg(p)=n=\sum_{j=1}^{k+1} i_j, \, \deg(q)=m=\sum_{j=1}^{k} i_j
\end{array}
$$
satisfy the bilinear hypergeometric equation (\ref{hypergeometric}) 
with $\lambda_{nm}$ given in proposition \ref{bilinears}.
\end{proposition}

$ {\cal W}[\psi_1(z)...\psi_k(z)]=\det||d\psi_i(z)/dz^j|| $ 
in (\ref{darboux})  denotes the Wronskian determinant.

\vspace{.2in}

To prove the proposition we need the following lemma by Crum \cite{Cru}
\begin{lemma}
Let $L$ be a given second order Sturm-Liouville operator
$$
L=\frac{d^2}{d\phi^2}+ u_0(\phi)
$$
with a sufficiently smooth potential $ u_0 $, 
and let $\{\psi_1,...\psi_k\}$ be 
its eigenfunctions corresponding to arbitrary
fixed pairwise different eigenvalues 
$\{\lambda_1,...\lambda_k\}$, i.e. $\psi_i\in \ker(L+\lambda_i)$, 
$i=1..k$. Then, for arbitrary
$\psi\in\ker(L+\lambda)$ the function
$$
\tilde \psi=\frac{{\cal W}[\psi_1...\psi_k,\psi]}{{\cal W}[\psi_1...\psi_k]}
$$
satisfies the differential equation
$$
\left(\frac{d^2}{d\phi^2}+{u_k}(\phi)+\lambda\right)\tilde\psi=0
$$
with
$$
{u_k}=u_0+2\frac{d^2}{d\phi^2}\ln{\cal W}[\psi_1...\psi_k]
$$
\end{lemma}

Proof of proposition \ref{equilibrium}:


Changing variables as in (\ref{new}) and making a gauge transformation
$$
L\to L_0=\nu L \nu^{-1}, \quad \frac{d}{d\phi}\ln\nu=\frac{U}{2\sqrt{P}}
$$
we get a formally self-adjoint operator
$$
L_0=\left(\frac{d}{d\phi}+\frac{U}{2\sqrt{P}}\right)\left(\frac{d}{d\phi}-\frac{U}{2\sqrt{P}}\right)=\frac{d^2}{d\phi^2}+u_0
$$
with eigenfunctions
$$
\psi_i=\nu Q_i, \quad (L_0+\lambda_i)\psi_i=0, \quad i=0,1,2,...
$$
According to the Crum lemma the function
\begin{equation}
\frac{{\cal W}[\psi_{i_1}(\phi),\psi_{i_2}(\phi),...\psi_{i_k}(\phi),\psi_{i_{k+1}}(\phi)]}{{\cal W}[\psi_{i_1}(\phi),\psi_{i_2}(\phi),...\psi_{i_k}(\phi)]}=\frac{\nu p}{q}
\label{tequi}
\end{equation}
is an eigenfunction of
$$
L_k=\frac{d^2}{d\phi^2}+u_k, \quad u_k=u_0+2\frac{d^2}{d\phi^2}\log{\cal W}[\psi_{i_1},\psi_{i_2},...\psi_{i_k}]=u_0-2\frac{d^2}{d\phi^2}\log(\nu^kq)
$$
with the eigenvalue $\lambda_{i_{k+1}}$. 

Deriving (\ref{tequi}) and the last equation we used the 
following properties of Wronskians
$$
{\cal W}[\nu f_1,...\nu f_n]=\nu^n{\cal W}[f_1,... f_n],\quad {\cal W}[f_1(z(\phi))...f_n(z(\phi))]=\left(\frac{dz}{d\phi}\right)^{\frac{1}{2}n(n-1)}{\cal W}[f_1(z)...f_n(z)]
$$
It then follows immediately that
\begin{equation}
\begin{array}{l}
\left(L_k+\lambda_{i_k+1}\right)\left[\frac{\nu p}{q}\right]=0=\\
\\
\quad\frac{1}{\nu q^2}\left(q\frac{d^2p}{d\phi^2}-2\frac{dq}{d\phi}\frac{dp}{d\phi}+p\frac{d^2q}{d\phi^2}+\frac{U}{ \sqrt{P}}\left(\frac{dp}{d\phi}q-\frac{dq}{d\phi}p\right)+(\Lambda_k+\lambda_{i_{k+1}})pq\right)
\end{array}
\label{inequ}
\end{equation}
where
$$
\Lambda_k=\frac{1}{2}k\left(2\frac{dU}{dz}-\frac{U}{P}\frac{dP}{dz}\right)
$$
It is clear from the statement of the proposition that 
$\Lambda_k$ is independent of $z$.

Finally, changing the independent variable back to $z$ (\ref{new}), 
$\phi=\phi(z)$ , we arrive to the bilinear hypergeometric operator in the
rhs of (\ref{inequ}). 
The degrees of $p$ and $q$ can then be evaluated 
from the highest powers of 
Wronskians (\ref{darboux}).

\vspace{.2in}

{\bf Example:} Consider, for 
instance, the equilibrium configuration corresponding to the sequence
\begin{equation}
I=2,4,6
\label{example}
\end{equation}
in the system with
$$
P=1,\quad U=-2z
$$
The eigenstates of the linear problem are Hermite polynomials $H_n(z)$
(see figure \ref{table}). Computing $p$ and $q$, with the help of (\ref{darboux}),
we obtain
$$
p=8192z^3(8z^6-12z^4+18z^2-15), \quad q=32z(4z^4+3-4z^2)
$$
Polynomial $p$ has a multiple root $z=0$ and this is a common root
with polynomial $q$.

Excluding common factors we have
$$
{\bar p}=256z^2(8z^6-12z^4+18z^2-15),\quad {\bar q}=4z^4-4z^2+3
$$
It can be verified without much difficulty that ${\bar q}$ and ${\bar p} $
do not have multiple roots, other than $z=0$. Hence sequence (\ref{example})
gives the following equilibrium distribution of charges 
( interacting via logarithmic potentials ) , 
in the linear external field: One charge of the value 
$\nu_1=+2$ at $z=0$,
six charges of the value $\nu_{2..7}=+1$ on the real line and 
four negative charges $\sigma_{1..4}=-1$
in the complex plane.

\vspace{.2in}

The following proposition is an analog of proposition 
\ref{equilibrium} for $P(z)=z$.
\begin{proposition}
\label{equilibri}
Let $I=i_1<i_2 ... i_k<i_{k+1}$ be a strictly increasing sequence of
nonegative integers and $k=0 \bmod 4$. Let $Q_i(z)=L_i^{(-1)}(-bz)$ 
be Laguerre polynomials satisfying
the hypergeometric equation
$$
(L+\lambda_i)Q_i(z)=0,\quad L=z\frac{d^2}{dz^2}+bz\frac{d}{dz}, \,\, b \not = 0
$$
Then polynomials $p$ and $q$
\begin{equation}
\begin{array}{ll}
p(z)=z^{\frac{1}{4}k(k+1)}{\cal W}[Q_{i_1}(z),Q_{i_2}(z), ...Q_{i_k}(z),Q_{i_{k+1}}(z)],\\
\\
q(z)=z^{\frac{1}{4}(k-1)k}{\cal W}[Q_{i_1}(z),Q_{i_2}(z), ...Q_{i_k}(z)] 
\end{array}
\label{laguerre}
\end{equation}
$$
\deg(p)=n=\sum_{j=1}^{k+1} i_j-\frac{1}{4}k(k+1), \,\, \deg(q)=m=\sum_{j=1}^{k} i_j-\frac{1}{4}k(k-1)\\ \\
$$
satisfy bilinear hypergeometric equation (\ref{hypergeometric}) 
with $P(z)=z$ and $U(z)=bz$.
\end{proposition} 

Proof: repeats the proof of proposition \ref{equilibrium}, 
except that now $k$ must be multiple of 4 in order to
(\ref{laguerre}) be polynomials.

\section*{VI. Rational and Trigonometric solutions of KP/KdV hierarchies,
Evolution in Two Dimensions}

Another interesting set of examples are limits $U=0$. 
They correspond to decreasing at infinity rational  
or periodic soliton solutions of the KP/KdV hierarchies 
(see equation (\ref{solitons}) ).

For instance, studying the case
$$
P=1, \quad U=0
$$
Bartman \cite{Bartman} provided an electrostatic 
interpretation for the Adler-Moser polynomials.

Indeed, in this limit
the bilinear hypergeometric equation becomes 
the recurrence relation for the Adler-Moser polynomials
\begin{equation}
p^{\prime\prime}q-2p^\prime q^\prime+pq^{\prime\prime}=0
\label{free}
\end{equation}
which, as shown by Burchnall and Chaundy 
(who, by the author knowledge, first 
studied (\ref{free}) in \cite{BC}), exhaust all
polynomial solutions of (\ref{free}).

Note that, in difference from the generic cases 
shown in the figure \ref{table}, we have a 
set of polynomials depending continuously
on $k+1$ parameters:
$$
p=\theta_{k+1}, \quad q=\theta_k, \quad \theta_k=\theta_k(z+t_1,t_2,...t_k)
$$
$$
\theta_k={\cal W}[\psi_1,...\psi_k], \quad \psi_j^{\prime\prime}=\psi_{j-1}, \quad \psi_0=1,  \psi_1=z, \quad \deg(\theta_k(z))=(k+1)k/2
$$
with the second logarithmic derivatives of $\theta$s being rational 
solutions of the KdV hierarchy.

Thus, we have (at generic values of $t_i$) 
equilibrium of $k(k+1)/2$ positive 
and $\frac{1}{2}(k+1)(k+2)$ negative free charges
with positions in the complex plane continuously depending on $t_i$.

\vspace{.15in}

Let us turn now to the following problem: Find homogeneous 
polynomials $p(X,Y,t),\, \deg(p)=n$ , $ q(X,Y,t),\, \deg(q)=m $ 
in two variables $X,Y$
satisfying the equation
\begin{equation}
\frac{dp}{dt}q-\frac{dq}{dt}p=(X^2+Y^2)\left(q\Delta p - 2( \nabla q , \nabla p ) + p\Delta q\right),
\label{homogenuous}
\end{equation}
where
$$
\Delta:=\frac{\partial^2}{\partial X}+\frac{\partial^2}{\partial Y}, \quad \nabla:=\frac{\partial}{\partial X},\frac{\partial}{\partial Y}
$$
and $(,)$ stands for the standard scalar product in ${\Bbb C}^2$.

Factorizing $p$ and $q$ as
$$
p=\prod_{i=1}^n(X\sin\phi_i-Y\cos\phi_i),\, q=\prod_{i=1}^m(X\sin\theta_i-Y\cos\theta_i)
$$
we come to the following

\begin{proposition}
The bilinear evolution (\ref{homogenuous}) induces dynamics
\begin{equation}
\begin{array}{l}
\frac{d\phi_i}{dt}=-2\sum_{ j \in N \not = i }\cot(\phi_i-\phi_j)+2\sum_{ j \in M }\cot(\phi_i-\theta_j)\\
\\
\frac{d\theta_i}{dt}=2\sum_{ j \in M \not = i }\cot(\theta_i-\theta_j)-2\sum_{ j \in N }\cot(\theta_i-\phi_j)
\end{array}
\label{quasitwo}
\end{equation}
\end{proposition}

Proof: It is convenient to write (\ref{homogenuous}) 
in the polar coordinates
$ (X=r\cos\phi,Y=r\sin\phi) $  
$$
p=r^n{\tilde p}=r^n\prod_{i=1}^n\sin(\phi-\phi_i), q=r^m{\tilde q}=r^m\prod_{i=1}^m\sin(\phi-\theta_i)
$$
\begin{equation}
{\tilde q}\frac{d\tilde p}{dt}-{\tilde p}\frac{d\tilde q}{dt}=\tilde q\frac{\partial^2\tilde p}{\partial\phi^2}-2\frac{\partial\tilde q}{\partial\phi}\frac{\partial\tilde p}{\partial\phi}+\frac{\partial^2\tilde q}{\partial\phi^2}{\tilde p}+(n-m)^2{\tilde p}{\tilde q}
\label{trigonometric}
\end{equation}

Then, it can be verified that equation (\ref{trigonometric}) 
corresponds to the case $P(z)=-z^2,\, U(z)=0$ 
in the classification of figure \ref{table}. 
It must be remarked, however, 
that solutions $\tilde p$ and $\tilde q$ are 
not polynomial, but algebraic functions of $z=\exp(2i\phi)$:
$$
\begin{array}{l}
{\tilde p}=z^{-n/2}\prod_{ j \in N }x_j^{-1/2}{\bar p}, \quad x_j=\exp(2i\phi_j),\, {\bar p}=\prod_{ j \in N }(z-x_j)\\ \\
{\tilde q}=z^{-m/2}\prod_{ j \in M }y_j^{-1/2}{\bar q},\quad y_j=\exp(2i\theta_j),\,{\bar q}=\prod_{ j \in M }(z-y_j)
\end{array}
$$
Nevertheless, since $\tilde p$ and $\tilde q$ are 
of ``almost'' polynomial type, we get (\ref{quasitwo}) by arguments 
similar to the proof of the proposition \ref{bilinears}. 
More precisely, for this purpose it is rather more convenient to 
use (\ref{lianerize}), where we can replace $q$ and $p$ with 
``pure'' polynomials ${\bar p}$ and ${\bar q}$ 
(and $\psi=\frac{p}{q}$ with $\frac{{\bar p}}{{\bar q}}$) : 
This substitution adds constants to coefficients
in (\ref{lianerize}) and 
 rhs of the equation acquires the common factor 
$z^{\frac{1}{2}(n-m)}\exp(i\sum_{ j \in N } \phi_j-i\sum_{ j \in M } \theta_j)$. 
The lhs of (\ref{lianerize}) acquires the same factor, 
since the quantity (``center of mass of the system'') 
$ \sum_{ j \in N } \phi_j-\sum_{ j \in M } \theta_j  $ 
does not change with time. The later statement can be easily 
verified adding equations of motion for $\phi$s and subtracting 
equations of motion for $\theta$s in (\ref{trigonometric}). 
Thus, the problem is reduced to the purely polynomial dynamics, 
which completes the proof.

The flow (\ref{quasitwo}) is a trajectory of two Sutherland systems 
in the absence of the external potentials.

It is interesting that the equilibrium condition 
for (\ref{trigonometric}) written in coordinates $X,Y$
\begin{equation}
q\Delta p - 2( \nabla q , \nabla p ) + p\Delta q=0
\label{laplace}
\end{equation}
has been studied in \cite{BeLou}, \cite{B} in connection with the 
Hadamard problem in the Minkowski space: Solutions of (\ref{laplace}) (or fixed points of (\ref{quasitwo}))
define differential operators possessing Huygens property in 
the Hadamard sense \cite{UMN}. The angular parts  
$  \tilde p , \tilde q  $ of solutions to (\ref{laplace}) are
periodic soliton solutions of the Korteveg-de Vries equation.

The following proposition provides us with $k+1$-parametric family 
of solutions to (\ref{laplace}) describing equilibrium configurations 
on the Coulomb charges (vortices) on the cylinder.

\begin{proposition}
\label{huygens}
Let $I=i_1<i_2...i_k<i_{k+1}$ be a strictly increasing sequence of 
nonnegative integers.
Then
$$
p=r^n{\cal W}[\psi_{i_1},\psi_{i_2},...\psi_{i_k},\psi_{i_{k+1}}],\,
q=r^m{\cal W}[\psi_{i_1},\psi_{i_2},...\psi_{i_k}]
$$
where
$$
\psi_{i_j}:=\sin(i_j\phi+t_i), \, {\cal W}:=\det|| d\psi_i/d\phi^j ||,\quad m=\sum_{j=1}^k i_j, n=\sum_{j=1}^{k+1}i_j 
$$
satisfy (\ref{laplace})
\end{proposition}

Proof: repeats the proof of the proposition \ref{equilibrium} 
for (\ref{trigonometric})=0, except that now we have a 
superposition of the
Thebyshev trigonometric polynomials 
$\sin(j\phi)$ and $\cos(j\phi)$ \cite{S} instead of $Q_j$.

\vspace{.2in}

Thus, in contrast with the plane case we have
$k+1$ continuous parameters and $k+1$ integers
defining equilibrium configurations on the cylinder.
Also, in difference from the plane distributions, the equilibrium 
is possible not only for consecutive triangle powers
 of the Adler-Moser polynomials, but for any two values of 
partitions $n=\deg(p)=\sum_{j=1}^{k+1} i_j$ and $m=\deg(q)=\sum_{j=1}^{k}i_j$.
This is due to a different topology of the problem:
roughly speaking, the charges on the cylinder have less 
"possibilities" to "escape" to infinity then on the plane. 
It must, however, be restated that
numbers and values of charges depend on multiplicities and common 
factors of $p$ and $q$.

\section*{VII. Polylinear Evolution Equations and Related Hamiltonian Systems}

As was mentioned in the introduction, (\ref{bieqs}) is, in fact, 
a special case of more general polylinear equation. 
The polylinear equation induces a polynomial dynamics which can be 
also embedded in a Hamiltonian flow. 
However, this flow does not separate now in independent components. 
The Hamiltonians are of the Calogero-Moser type for several species 
of interacting particles. They are not generally related to the 
Coxeter reflection groups.

Let us begin by introducing $l$ species of particles with distinct 
charges $Q:=\{Q_i,i=1..l\}$, $Q_i\not=Q_j, i \not = j $. 
We define the $l$-linear differential operator operator
\begin{equation}
\begin{array}{r}
H_\lambda^Q[q_1,...q_l](z)=P(z)\left(\sum_{i=1}^lQ_i^2q^{\prime\prime}_i(z)\prod_{n\not=i}^lq_n(z)+2\sum_{i<j}^lQ_iQ_jq_i^\prime(z) q_j^\prime(z)\prod_{n\not=i\not=j}^lq_n(z)\right)\\
\\
+\frac{1}{2}P^\prime(z)\sum_{i=1}^lQ_i^2q^\prime_i(z)\prod_{n\not=i}^lq_n(z)+U(z)\sum_{i=1}^lQ_iq^\prime_i(z)\prod_{n\not=i}^lq_n(z)+\lambda\prod_{i=1}^lq_i(z)
\end{array}
\label{manyq}
\end{equation}
acting on polynomials $ q_i \in  V_i \cong { \Bbb C } ^ { n_i + 1 } , i = 1 . . l $
\begin{equation}
q_i(z)=\prod_{j=N_i+1}^{N_i+n_i}(z-z_j(t)), \quad N_i=\sum_{j=1}^{i-1} n_j
\label{polyns}
\end{equation}
For convenience, we now use unique numeration for roots of all polynomials. 

Similarly to the case $ l =2, Q_1=1, Q_2=-1 $ of the bilinear 
hypergeometric operator (\ref{bihop}), the following 
proposition holds (We skip proofs below, since they repeat arguments 
of preceding sections).

\begin{proposition}
\label{many}
The polylinear operator  $H^Q_{\lambda_{n_1,..,n_l}}: 
V_1\times V_2\times...\times V_l \to V_{l+1} $, 
$ V_{ l +1 } \cong { \Bbb C } ^ { \sum_{i=1}^l n_i  } $
given by (\ref{manyq}) with
$$
P(z)=A+Bz+Cz^2,\quad U(z)=a+bz,
$$
and
$$
\lambda_{n_1,..,n_l}=-\left(U^\prime+\frac{1}{2}P^{\prime\prime}\sum_{i=1}^lQ_in_i\right)\sum_{i=1}^lQ_in_i
$$
induces dynamics
\begin{equation}
\frac{dz_i}{dt}=-2P(z_i)\sum_{j\not=i}\frac{Q_j}{z_i-z_j}-U(z_i)-\frac{Q_i}{2}P^\prime(z_i)
\label{miyyn}
\end{equation}
by action
$$
\sum_{i=1}^lQ_i\frac{dq_i}{dt}\prod_{n\not=i}^lq_n=H^Q_{\lambda_{n_1,...,n_l}}[q_1,..,q_l]
$$
on (\ref{polyns}).
\end{proposition}
Under conditions mentioned before, (\ref{miyyn}) 
describes motion/equilibrium
of $n_1$ charges $Q_1$, $n_2$ charges $Q_2$ etc.

To avoid confusion, we note that in (\ref{miyyn}) and up to 
the end of this section, the summation indexes go from $1$ 
to the total number of roots $\sum_{ j=1  } ^ l n_j$ and to 
each root $z_j$ we assign $Q_j$ which is equal to the charge of the 
corresponding polynomial.

It is natural to ask the following important question: May dynamics (\ref{miyyn}) be embedded
in a Hamiltonian flow?  We address it in the following 

\begin{theorem}
\label{hamito}
(\ref{miyyn}) is a trajectory of the Hamiltonian system
\begin{equation}
\label{newsyst}
H=\sum_j\frac{Q_j}{2P(z_j)}\left(\frac{dz_j}{dt}\right)^2-\sum_{j}Q_j\frac{U_{Q_j}(z_j)^2}{2P(z_j)}-\sum_{k<j}Q_kQ_j(Q_k+Q_j)W(z_k,z_j)
\end{equation}
where
$$
W(z_1,z_2)=\frac{P(z_1)+P(z_2)}{(z_1-z_2)^2}, \quad U_{Q_j}(z)=U(z)+Q_jP^\prime(z)/2
$$
\end{theorem}
System (\ref{newsyst}) is of the Calogero(Sutherland)-Moser 
type of $l$ species of particles with masses $Q_i$. 
In this picture, the two-body potentials are of the similar form, 
but having different amplitudes $Q_jQ_k(Q_j+Q_k)$ 
(for interaction within each of the species of particles and between 
the species respectively). They are translation invariant in 
Newtonian coordinates (\ref{new}) if (up to a linear transformation) 
$P(z)=1$ or $z^2$. We remind that in the case $ l = 2, Q_1=1, Q_2=-1 
$ the interaction between two different species vanishes, leading to 
separation of the Hamiltonians, while for $l=1$ we obtain identical 
Calogero-Moser particles. Both above cases are related to $A/BC/D$ 
root systems.

It is seen without much difficulty that for generic $l,Q$, system 
(\ref{newsyst}) is not related to any Coxeter reflection group. 
Although , by the author knowledge, quantum models related to 
different deformations of the Coxeter root systems were considered 
in earlier works ( eg \cite {BeLou}, \cite{V} ), (\ref{newsyst}) 
has not appeared in the literature. We
do not attempt to address the question of integrability of 
(\ref{newsyst}) in this paper, leaving it for future studies. 

\section*{VIII  General Bilinear Dynamics, Evidences of Integrability}

Let us concentrate on the two-component case $l=2, Q={1,-\Lambda}$ 
of general
bilinear hypergeometric operator (\ref{smile}),
restricting ourselves with the rational case
$$
P(z)=i, U(z)=i\omega z, \quad i:=\sqrt{-1}
$$
Such a choice of the coefficients leads to dynamical system
\begin{equation}
\begin{array}{l}
i\frac{d}{dt}x_j=2
\left(
\sum_{k \in N \not=j}\frac{1}{x_k-x_j}-\sum_{k\in  M}\frac{\Lambda}{x_j-y_k}
\right)
+\omega x_j, \, j \in N\\
\\
i\frac{d}{dt}y_j=2
\left(
-\sum_{k \in M \not=j}\frac{\Lambda}{y_j-y_k}+\sum_{k \in N}\frac{1}{y_j-x_k}
\right)
+\omega y_j, \, j\in M
\end{array}
\label{biyyn}
\end{equation}
of $n$ and $m$ particles of two different types. 
According to proposition \ref{many} and theorem \ref{hamito}, 
(\ref{biyyn}) is a corollary of the evolution equation
\begin{equation}
iq\frac{dp}{dt}-i\Lambda p\frac{dq}{dt}=p^{\prime\prime}q-2\Lambda p^\prime q^\prime+\Lambda^2pq^{\prime\prime}
+\omega z\left(p^\prime q-\Lambda pq^\prime\right)+\omega (\Lambda m-n) pq
\label{nones}
\end{equation}
for polynomials $p$ and $q$ (\ref{polyn}), and  is a trajectory 
of a system with the Hamiltonian
\begin{equation}
H=\frac{1}{2}\sum_{ j \in N }\left(\left(\frac{dx_j}{dt}\right)^2+\omega^2 x_j^2\right)-\frac{\Lambda}{2}\sum_{ j \in M }\left(\left(\frac{dy_j}{dt}\right)^2+\omega^2 y_j^2\right)+V(x,y)
\label{integrable}
\end{equation}
$$
V(x,y)=\sum_{ k < j \in N  }\frac{2}{(x_j-x_k)^2}+\sum_{ j \in N , k \in M  }\frac{\Lambda(\Lambda-1)}{(x_j-y_k)^2}-\sum_{ k < j \in M  }\frac{2\Lambda^3}{(y_j-y_k)^2}
$$ 
{\bf Remark}: Although, similarly to the case $\Lambda=1$, (\ref{nones}) can be 
written in the form of a time dependent Schr\"odinger equation (\ref{lianerize}), (changing $t$ to $it$) 
with the ``$\psi$'' and ``$\tau$'' functions given by 
$$
\psi=p/q^\Lambda,\tau=q^{\frac{1}{2}\Lambda(\Lambda-1)}
$$
the dynamics of poles of the potential ${\cal U}=2(\log q)^{\prime\prime}$
cannot be embedded in a Hamiltonian flow uncoupled from the dynamics 
of zeros of $p$. This is why system (\ref{biyyn}) cannot be connected with solutions of
 the KP hierarchy.

\vspace{.25in}

Although the Hamiltonian system (\ref{integrable}) unlikely 
be integrable for arbitrary initial conditions and 
$\Lambda\not=0,\pm1$, we find that its trajectories (\ref{biyyn}) 
(defined by the polynomial evolution (\ref{nones})) are integrable.
\begin{conjecture}
\label{intel}
System (\ref{biyyn}) is completely integrable for arbitrary real 
$\Lambda$ and $\omega$ in the sense that there exist $2(n+m)-1$ 
functionally independent integrals of motions $I_j$, 
which are real rational functions of $x,y$, i.e.
$I_j=I_j(x_1,x_1^*,...x_n,x^*_n,y_1,y^*_1,..,y_n,y^*_n)=I_j(x^*_1,x_1,..,x_n^*,x_n,y^*_1,y_1,..,y^*_n,y_n)$, $j=1..2(m+n)-1$
\end{conjecture}

\begin{figure}
\begin{center}
\leavevmode
\epsfxsize=368pt
\epsfysize=390pt
\epsfbox{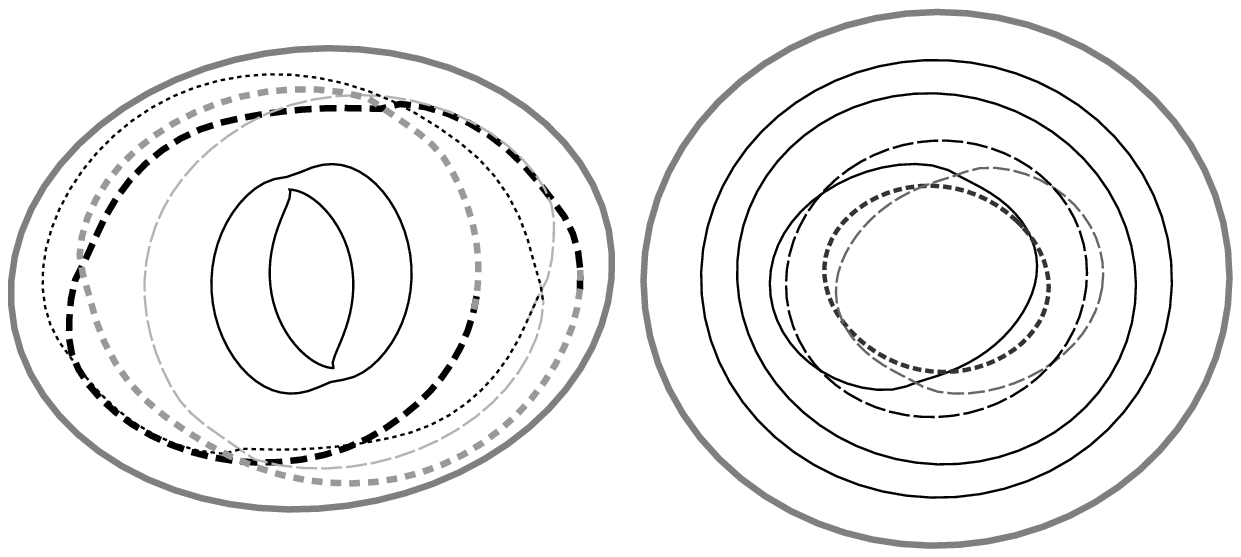}
\end{center}
\caption{Examples of trajectories of the two component system 
consisting of $n=6$ unit charges and $m=1$ charge $-\Lambda$, 
where $\Lambda=1.213579$. Every curve shows an individual trajectory 
of each charge in its own coordinates $(\Re x_j, \Im x_j), j=1..6$ or 
$(\Re y_j, \Im y_j), j=1$. The charge of value $-\Lambda$ is depicted by the 
gray solid line. The motion shown on the left figure has period 
$4\pi/\omega=2T$, while the period on the right figure is equal to 
$T=2\pi/\omega$. The motion on the left is depicted within the time interval 
$t=[0,T]$, which is a half-period for such initial conditions. 
In this case trajectories of several charges coincide interchanging 
each half-period $T$.}
\label{turn}
\end{figure} 

We devote the rest of this section to examples in favor of this conjecture.

We take the case $\Lambda=1$ as the first example: 
According to corollary \ref{only} the equations of motion can be 
reduced to the Lax form (see eg \cite{P})
$$
i\frac{d}{dt}L_x=[L_x,A_x]+\omega L_x, \, i\frac{d}{dt}L_y=[L_y,A_y]+\omega L_y
$$ 
where $L_y,A_y$ and $L_y,A_y$ are matrices of dimensions 
$n\times n, m\times m$ respectively
\begin{equation}
\begin{array}{c}
(L_x)_{jk}=\frac{1}{2}\left(i\frac{dx_j}{dt}+\omega x_j\right)\delta_{kj}+\frac{1-\delta_{jk}}{x_j-x_k},\quad j,k\in N \\
\\
(L_y)_{jk}=\frac{1}{2}\left(i\frac{dy_j}{dt}+\omega y_j\right)\delta_{kj}+\frac{1-\delta_{jk}}{y_j-y_k},\quad j,k\in M
\end{array}
\label{inte}
\end{equation}
Substituting (\ref{biun}) to (\ref{inte}) we eliminate 
velocities $dx/dt, dy/dt$, getting the Lax matrices $\tilde {L_x}$ 
and $\tilde {L_y}$ for (\ref{biun}), depending on the coordinates 
only. It is easy to see that the absolute values of squares of traces
\begin{equation}
I_1=({\rm Tr} L)({\rm Tr} L )^* ,\quad I_2= ({\rm Tr} L^2)({ \rm Tr } L^2 )^* ,..,I_{2(n+m)-1}= ( { \rm Tr } L^{2(n+m)-1} ) ( { \rm Tr } L ^ { 2 ( n + m ) - 1 } )^*
\label{motion}
\end{equation}
of the $ (m+n) \times (m+n) $ matrix
$$
L=\left(\begin{array}{cc}{\tilde L_x}&0\\0&{\tilde L_y}\end{array}\right)
$$
are real rational integrals of motion. They are homogeneous 
functions in $x,y,\omega$ 
(with $x,y$ and $\omega$ having weights $-1,-1,2$ respectively). 
The functional independence of (\ref{motion}) can be 
easily proven by considering them as polynomials in $\omega$ with 
functionally independent highest symbols
$$
I_k=\omega^{2k}\left(\sum_{j\in N}x_j^k+\sum_{j\in M}y_j^k\right)\left(\sum_{j\in N} ( x^*_j )^k+\sum_{j\in M}( y^*_j )  ^k\right)+..,\quad   k=1..2(n+m) - 1
$$
Let us now turn to the general system $\Lambda\not=0,\pm1$. 
Although, in this case, our arguments in favor of 
inetgrability of (\ref{biyyn}) stem mainly from numerical studies, we would like to mention some analytic results:

The Hamiltonian (\ref{integrable}) admits total separation of variables in low dimensions $n+m<4$. Namely, separating motion of the center of mass, we obtain one or two dimensional problem, admitting (in the latter case) further separation of variables in the polar coordinates.

Another (less trivial) example is the system with even number of unit charges $n=2l$ and a single particle of the second type $m=1$ having an arbitrary charge $-\Lambda$. The system is subject to symmetric initial conditions:
$$
x_j(0)=x_{j+l}(0), \quad y_1(0)=0, \quad  j=1..l
$$ 
It is seen without much difficulty that due to this ${\Bbb Z}_2$ symmetry, the above conditions hold for any $t$. Taking this fact into account, we may reduce (\ref{biyyn}) by this symmetry keeping only variables $x_j, j=1..l$. Changing the variables as $ x_j=\sqrt{z_j} $ we arrive to the following equations of motion
\begin{equation}
i\frac{dz_j}{dt}=-\sum_{k\not=j}\frac{4z_j}{z_j-z_k}+\omega
\label{problem}
\end{equation}
which corresponds to the $ { \rm BC } _ n $ rational case of figure \ref{table}. The integrability of (\ref{problem}) is then provided by arguments used for the study of linear dynamics. The rational integrals of motion for (\ref{problem}) can be found using the Lax representation for the Calogero-Moser system of the $ { \rm BC } _ n $ type.

One can also prove the periodicity of small nonsymmetric deviations from the symmetric trajectories as linear perturbations around (\ref{problem}). We do not perform this analysis here, since it requires cumbersome calculations. 

Finally, numerical simulations show 
that for any initial conditions and real $\Lambda$ trajectories of 
(\ref{biyyn}) turn out to be periodic, which shows existence 
of $2(n+m)-1$ independent rational integrals of motion. 
The period of motion is 
an integer multiple the period of the ``free'' oscillator 
$T=2\pi/\omega$, with an integer factor depending on the initial conditions.

Typical examples of trajectories for generic 
initial conditions and $\Lambda$ are shown on figure \ref{turn}.

\section*{IX. Conclusions and Open Questions}

The results at which we have now arrived may be summed up as follows: 
The bilinear hypergeometric operator (\ref{bihop}) induces dynamics 
(\ref{bidyn}), which may be embedded in a Hamiltonian flow.
In the case $\Lambda=1$ this flow is generated by a sum of two 
independent Calogero(Sutherland)-Moser Hamiltonians 
(theorem \ref{hamiltonii}) with (generally) different forms
of external potentials. This allows us to integrate (\ref{biun}) by 
the Lax method. The fixed points of the bilinear evolution
correspond to equilibrium distributions of different species 
of point vortices on the plane or cylinder. They may be
obtained (again, for $\Lambda=1$) by a finite number of Darboux 
transformations from the eigenstates of associated linear problems (propositions \ref{equilibrium}, \ref{equilibri}, \ref{huygens}). The dynamical system (\ref{biyyn}) of two species of interacting points in an external field is conjectured to be completely integrable for arbitrary real $\Lambda$ and $\omega$.

Let us now mention some open questions

The main problem, of course, is to prove 
integrability of (\ref{biyyn}) for arbitrary 
$\Lambda$ (conjecture \ref{intel}). 
It might be done by using two approaches: The first approach is to
find a 
Lax representation for (\ref{biyyn}). The Lax matrices for arbitrary $\Lambda$ 
could have a complicated structure, being rational functions
of $x,y$ and $\omega$ of greater homogeneity degree in comparison 
with
 the Calogero-Moser case.
It makes 
hard to find them using straightforward computational approaches.

Another method is to try to linearize (\ref{nones}). Although, as was
mentioned before, such
a linearisation, connected with the KP equation, was possible for 
$P=1$ and $\Lambda=1$, we cannot apply a similar scheme in general
case.

Another set of questions is connected with solutions of bilinear 
hypergeometric equation:

In 1929 \cite{BC}, Burchenall and Chaundy studied the following 
question: What conditions
must be satisfied by two polynomials $ p(z), q(z) $ in order that the indefinite
integrals
$$
\int \left(\frac{p(x)}{q(x)}\right)^2dx,\quad \int\left(\frac{q(x)}{p(x)}\right)^2dx
$$
may be rational, provided $p$ and $q$ do not have multiple and 
common roots?

They found that the integrals are rational if $p=\theta_i$, 
$q=\theta_{i+1}$ are 
(now know as the Adler-Moser) polynomials satisfying (\ref{free}).

From the other hand, we know that any two polynomial solutions of 
the ordinary
hypergeometric equations $p=Q_n, q=Q_m$ are orthogonal with the 
measure $\nu(z)$
$$
\int \nu(z) Q_n(z) Q_m(z) dz = 0,  n\not=m
$$
Since the two above integrals are related to particular forms of (\ref{bihyp}),
it is natural to ask the following question: what integration condition
may be imposed on two polynomials $p$ and $q$ in order for they satisfy a nondegenarate bilinear hypergeometric equation (\ref{bihyp})?

In the same work \cite{BC}, Burcnall and Chaundy have shown 
that any polynomial solutions
of (\ref{free}) may be obtained by a finite number of Darboux 
transformation from the kernel of the
"free" differential operator $d^2/dz^2$.

In this paper we have proved proposition \ref{equilibrium}, stating
that polynomials obtained from the eigenfunctions of the ordinary 
hypergeometric equation by a finite number of Darboux transformations
are solutions of (\ref{bihyp}). By analogy with \cite{BC} it 
is natural to state the following

\begin{conjecture}
\label{niyrt}
Any polynomial solutions to bilinear hypergeometric equations of 
propositions \ref{equilibrium}, \ref{equilibri} are  
(\ref{darboux}) and (\ref{laguerre}) respectively.
\end{conjecture}

Concluding the article we would like to mention briefly possible 
multi-dimensional generalizations of
(\ref{bihyp}). A particular generalization was constructed
in section VI for the homogeneous polynomials in two variables 
(\ref{laplace}). A similar construction \cite{B} related to the
classical special functions in many dimensions \cite{HOP}, is 
connected with the quantum Calogero-Moser systems on the
Coxeter root systems (and their deformations \cite{BeLou} , 
\cite{V}). In this context, it would be interesting to find a 
proper analog of (\ref{bihyp}) in many dimensions.

\section*{Acknowlegements}

The author is grateful to H.Aref, F.Calogero, B.Dubrovin, A.Kirillov, A.Orlov for useful information and remarks.

\end{document}